\documentclass[12pt]{article}
\usepackage{graphicx}
\usepackage{amsmath}  

\begin{document}
\def\be{\begin{equation}}
\def\ee{\end{equation}}

\begin{center}
{\large\bf{ Determination of the refractive index of water and glass using smartphone cameras by estimating the apparent depth of an object}}
\end{center}

\begin{center}
{Sanjoy Kumar Pal$^1$,Soumen Sarkar$^2$, Surajit Chakrabarti$^3$ \\
$^1$Anandapur H.S. School, Anandapur,Paschim Medinipur,WB,India\\
$^2$Karui P.C.High School, Hooghly, WB, India\\
$^3$Ramakrishna Mission Vidyamandira, Belur Math, Howrah, WB,India\\}
E-mail: sanjoypal83@gmail.com;tosoumen84@gmail.com; surnamchakrabarti@gmail.com
\end{center}

\section{Abstract}
A smartphone camera can be used for measuring the width and distance of an object by taking its photographs. The focal length of the camera lens can be determined very accurately by finding the width of the image of an object on the camera sensor to micron level accuracy. The level of accuracy achieved with the help of camera sensors, allows us to determine the refractive index of water upto four significant digits by finding the apparent depth of an object immersed in it. We have also measured the refractive index of glass by the same method, using  three glass slides of different thicknesses, the smallest being 1.2 mm.

\section{Introduction}
A smart phone is a powerful instrument in the hands of a large section of people around the world. Students of physics in particular, can use it to  perform many experiments, as is evident from different publications in educational journals from all over the world. We have performed an experiment on optics using the cameras of smartphones. Camera sensors in smartphones use sophisticated technology to produce images in a two dimensional array, called pixels, with length dimensions accurate to microns. We  have taken the advantage of this accuracy to measure the refractive index of water and glass with great accuracy, by taking photographs of an object and its apparent positions when  kept beneath water or glass. 

In some recent works [1,2] the method of determining the focal length of the lens of a smartphone camera, has been discussed. In the present paper we have first determined the focal lengths of lenses in two smartphone cameras  used in this experiment. By photographing an object from a suitable position, we can find the image size on the camera sensor in terms of pixel, using software called Paint and Preview, available freely  from the internet for Windows and Apple Mac operating systems, respectively. If we know the transverse dimensions of the object, we can easily find its transverse magnification. From the magnification and the focal length of the lens, we can accurately estimate the distance of the object[3]. This allows us to find the apparent depth of an object placed beneath  water or  glass, very accurately. From this we determine the refractive index of water or glass by the standard formula [4-9] \be \mu=\frac{\text{Real Depth}}{\text{Apparent Depth}}.\ee We have measured the refractive indices  of water and glass with  Apple smartphones  of model  IPhone 12-Mini and IPhone 12 Pro Max respectively.

\section{Determination of the  focal length of the lens}

We can easily show that the object distance $u$ from a convex lens of focal length $f_c$ can be related to its transverse magnification $m$ as [3] \be u=f_c(\frac{1}{m}-1).\ee In our sign convention, the focal length is positive and the object distance for the real image is  negative and the magnification is negative. If we know the object size and the focal length of the lens, we can find the distance of the object using the algebraic equation (2). 

For an accurate determination of the focal length of the lens, we photograph the object from two positions of the camera shifted by a distance $D$ along the line of sight of the lens.
If $u1$ and $u2$ are the  distances of the  object from two positions of the camera  with transverse magnifications $m_1$ and $m_2$ respectively, we get 
\be D=f_c(\frac{1}{m_2}-\frac{1}{m_1})\ee where $D=u2-u1$.This displacement should be along the line of sight of the camera lens. In terms of the object and image sizes we can write $m_1=\frac{I_1}{O}$ and $m_2=\frac{I_2}{O}$ where $O$ is the size of the object and $I_1$ and $I_2$ are the sizes of the images formed on the camera sensor. We can write equation (3) as
\be f_c=\frac{\left |D\right |}{\left |(\frac{1}{m_2}-\frac{1}{m_1}\right)|}.\ee
In equation (3) the displacement $D$ can be both towards or away from the object photographed. In figure 1, we show the lens configuration. The lens in the displaced position has been shown with dash marks. In this figure $D$ is positive and $m_2$ is less than $m_1$. So, $f_c$ becomes positive as it should. Similarly, when the camera is brought closer to the object, $D$ is negative and the denominator in equation (3) is also negative. So, we have written equation (4) with absolute signs. We determine $f_c$ using equation (4) by determining $I_1$, $I_2$ of an object of known width $O$ and known displacement $D$ of the lens. Once we determine the focal length, we can determine the distance of an object using equation (2). 
\section{Experiment}
We first determine the focal lengths of the two camera lenses. In table 1 we present the data for one lens. We take a plastic ruler and photograph it from two distances with displacement $D$ along the line of sight of the lens. $u1$ and $u2$ are the two distances  of the camera from the ruler. When the image of the ruler is opened by the software Paint, one can select a certain part of the image using the cursors of the software and can find the pixel difference between its two ends. We select 7cm equivalent length on the image plane and the pixel equivalent has been presented as the pixel diff. in the table. From the internet one can find the pixel length for the particular model of the camera. For our lens, 1 pixel was equal to 1.4 $\mu m$ [10]. From this we determine the magnifications of the images which are obtained by taking the ratio of the image size to the object size which was 7cm.  From this and the displacement of the camera, we determine the focal length very accurately using equation (3). There is an uncertainity in the actual position of the lens inside the smartphone. We displace the smartphone by a known distance in order to eliminate this uncertainity. Here the displacement of the lens is the same as the displacement of the phone. In exactly the same way, we determine the focal length of the other camera lens for which 1 pixel is equal to 1.7 $\mu m$ [10].

To determine the refractive index of water we take an empty bucket and keep a small ruler at the bottom. From a distance about 10cm above the bucket we photograph the ruler with a smartphone held by a stand. By analysing the photograph with the Paint software, we can easily find  the image size in terms of pixels. Specifically we find the image width for two marks on the ruler separated by 4cm and find the magnification. Using equation (2) we get the distance of the ruler from the camera. Now putting some water in the bucket we can, in the same way, find the distance of the virtual object. Finally, we float  the ruler on the surface of water in the bucket and hence find the distance of the surface of water from the camera. All these three photographs are taken from the same position of the camera. From these three measurements we can determine the real depth as well as the apparent depth of the object. We then determine the refractive index of water using equation (1). 

In a similar way, we determine the refractive index of glass. We take three  slides of the same glass material, of nominal thicknesses of 1,4,6 mm. We take a strip of a graph sheet secured on a wooden stool. We take a  photograph of the graph sheet  from a fixed distance with a smartphone held by a stand. Then we put the glass slide, of a definite thickness, on the graph sheet. We take the photograph of the graph paper through the glass slide. Finally we remove the graph paper from the bottom of the glass slide and secure it on top of the glass. Then we take the third photograph of the graph sheet. Here also the camera position is kept fixed for the three photographs. From these photographs we determine the real depth  which is nothing but the thickness of the glass slide. For each glass slide we determine the apparent depth inside the glass. From these we determine the refractive index of glass.

\section{Experimental Results}
The first row of table 1 shows that our camera is at a distance of $u1=30.0 $cm from the ruler. The camera was then shifted by 100.0cm to take the second photograpph from a distance $u2=130.0$cm. Magnifications in these two positions are determined and using equation (4) we determine the focal length of the lens. The  process is repeated for 9 more times varying $u1$ and $D$.
  The average focal length turns out to be $f_c=0.423\pm 0.001$cm. Similarly we get the focal length of the second camera lens $f_c=0.531\pm 0.001$cm. We have not shown the data for this camera in tabular form.
  
  In tables 2 and 3 we discuss the measurement of the refractive index of water. In  table 2, $u1$ denotes the distance of the ruler, placed at the bottom of the empty bucket, from the camera. The bucket is  partially filled with water. The distance of the apparent position of the ruler, as
  seen by the camera, is $u3$.  After taking two photographs we take up the ruler and  float it carefully on the water. Once again it is photographed. Its distance from the camera is denoted by $u2$. Each row of table 2 corresponds to the same row of table 3.Image sizes are determined from pixel readings and the transverse magnifications are obtained. Using equation (2) we calculate all distances $u1$, $u2$, and $u3$. From these we determine the real depth and the apparent depth of the ruler and hence the refractive index. We get the average value of the refractive index of water as $\mu=1.327\pm 0.004$. The refractive index of pure water with respect to air at $20^{\circ}$C has been given by [11] as 1.3312 at 656 nm. Our sample of water is not pure. The refractive index of water  goes down with the rise in temperature [12]. In our case the room temperature was $30^{\circ}$C. Assuming the presence of a small amount of dissolved material in our sample, we can claim that our result upto 4 significant figures matches that given in [11] within experimental uncertainty. 
  
We have determined the refractive index of glass using a second smartphone camera. In tables 4 and 5 we show our data with the three glass slides. Here also each row of table 4 corresponds to the corresponding row of table 5. For each slide we determine $u1$,$u2$,$u3$ from the same position of the camera in the same way as described for water. For each slide we have taken photographs fron 4 positions. As shown in table 4 the average measured thicknesses of the slides turn out to be 1.2 mm, 4.0 mm and 6.0 mm. In table 5 we show the average refractive index of glass slides which we determine to be  $1.50\pm 0.01$. 

\section{Conclusions}

Taking advantage of the accuracy to micron level in the image sizes on the sensor of smartphone cameras, we have been able to determine the refractive index of water to 4 significant digits. We have measured the apparent depth of an object immersed in water by  taking photograph of the object, looking vertically downwards. Using the same technique we have determined the refractive index of glass taking glass slides with thickness as small as 1.2 mm. The accuracy of the refractive index also indicates that our measurement of the thickness of the glass slides is also accurate. So the photographic technique introduced here is an alternative method for determination of the thickness of a thin sheet.The technique  is easy and fast. We can determine the refractive index of any transparent liquid or a transparent solid, in the shape of slides, very accurately, using this photographic method.

\begin{table}[ht]
\centering
\caption{Pixel data for the image sizes on the camera sensor from two distances:\\Object size = 7cm\\Smartphone :Apple IPhone 12-Mini
; 1 pixel=1.4$\mu m$}
\begin{tabular}{cccccccc}
\hline
obs&$u_1$&pixel&$I_1$&$u_2$&pixel &$I_2$&$f_c$\\
no.&cm&diff. 1&cm&cm&diff. 2&cm&cm\\
\hline
1&30.0&712&0.0997&130.0&163&0.0228&0.422\\
2&40.0&537&0.0752&140.0&152&0.0213&0.424\\
3&30.0&712&0.0997&140.0&152&0.0213&0.426\\
4&40.0&537&0.0752&150.0&142&0.0199&0.425\\
5&50.0&427&0.0598&160.0&132&0.0185&0.421\\
6&40.0&537&0.0752&160.0&132&0.0185&0.421\\
7&50.0&427&0.0598&170.0&124&0.0174&0.421\\
8&30.0&712&0.0997&160.0&132&0.0185&0.422\\
9&40.0&537&0.0752&180.0&118&0.0165&0.423\\
10&30.0&712&0.0997&180.0&118&0.0165&0.424\\

\hline
\end{tabular}
\end{table}

\begin{table}[ht]
\centering
\caption{Data for refractive index of water\\u1=Distance to the bottom of the bucket from the camera\\u2=Distance to the top of the water surface from the camera\\Object size:4cm\\Focal length of the lens $f_c$=0.423 cm}
\begin{tabular}{cccccccc}
\hline
obs &pixel&m1&u1&pixel&m2&u2&Depth of water\\
no.&diff. 1&&cm&diff. 2&&cm&(u1-u2)cm\\
\hline
1&303&0.01060&40.33&333&0.01166&36.70&3.63\\
2&240&0.00840&50.78&272&0.00952&44.86&5.92\\
3&303&0.01060&40.33&373&0.01306&32.81&7.52\\
4&316&0.01106&38.67&401&0.01404&30.55&8.12\\
5&329&0.01152&37.14&449&0.01572&27.33&9.81\\
6&329&0.01152&37.14&601&0.02104&20.53&16.61\\
7&329&0.01152&37.14&772&0.02702&16.08&21.06\\
8&329&0.01152&37.14&1065&0.03728&11.77&25.37\\

\hline
\end{tabular}
\end{table}

\begin{table}[ht]
\centering
\caption{Data for refractive index of water\\u3= Distance of the virtual object from the camera\\ Object size :4cm\\Focal length of the lens $f_c$=0.423cm}
\begin{tabular}{ccccccc}
\hline
obs &pixel&m3&u3&Apparent depth&$\mu=\frac{u1-u2}{u3-u2}$&Av $\mu$\\
no.&diff. 3&&cm&(u3-u2)cm&&\\
\hline
1&310&0.01085&39.41&2.71&1.339&\\
2&247&0.00864&49.38&4.52&1.310&\\
3&318&0.01113&38.43&5.62&1.338&\\
4&333&0.01166&36.70&6.15&1.320&$1.327\pm 0.004$\\
5&352&0.01232&34.76&7.43&1.320&\\
6&371&0.01298&33.01&12.48&1.331&\\
7&383&0.01340&31.99&15.91&1.324&\\
8&397&0.01390&30.85&19.08&1.330&\\

\hline
\end{tabular}
\end{table}

\begin{table}[ht]
\centering
\caption{Data for refractive index of glass\\Object size=6cm\\Smartphone:Apple IPhone 12 pro max\\Pixel size=1.7$\mu m$; Focal length of the camera lens $f_c=0.531 \pm 0.001$cm\\u1=Distance of the graph sheet from the camera\\u2=Distance to the top of glass slide}
\begin{tabular}{ccccccccc}
\hline
&Nominal thickness&pixel&m1&u1&pixel&m2&u2&Thickness of glass\\
obs&of slides&diff. 1&&&diff. 2&&&slides(u1-u2)\\
no.&mm&&&cm&&&cm&cm\\
\hline
1&&1749&0.04956&11.24&1770&0.05015&11.12&0.12\\
2&1&1606&0.04550&12.20&1623&0.04598&12.08&0.12\\
3&&1504&0.04261&12.99&1519&0.04304&12.87&0.12\\
4&&1321&0.03743&14.72&1333&0.03777&14.59&0.13\\
\hline
5&&1659&0.04700&11.83&1719&0.04870&11.43&0.40\\
6&4&1435&0.04066&13.59&1479&0.04190&13.20&0.39\\
7&&1282&0.03632&15.15&1317&0.03732&14.76&0.39\\
8&&1207&0.03420&16.06&1238&0.03508&15.67&0.39\\
\hline
9&&1659&0.04700&11.83&1751&0.04961&11.23&0.60\\
10&6&1435&0.04066&13.59&1503&0.04258&13.00&0.59\\
11&&1282&0.03632&15.15&1336&0.03785&14.56&0.59\\
12&&1207&0.03420&16.06&1255&0.03556&15.46&0.60\\
\hline
\end{tabular}
\end{table}

\begin{table}[ht]
\centering
\caption{Data for refractive index of glass\\u3= Distance of the virtual object from the camera\\ Object size :6cm}
\begin{tabular}{ccccccc}
\hline
obs &pixel&m3&u3&Apparent depth&$\mu=\frac{u1-u2}{u3-u2}$&Av $\mu$\\
no.&diff. 3&&cm&(u3-u2)cm&&\\
\hline
1&1756&0.04975&11.20&0.08&1.50&\\
2&1612&0.04567&12.16&0.08&1.50&\\
3&1509&0.04276&12.95&0.08&1.50&\\
4&1325&0.03754&14.68&0.09&1.44&\\
\hline
5&1679&0.04757&11.69&0.26&1.54&\\
6&1449&0.04106&13.46&0.26&1.50&$1.50\pm 0.01$\\
7&1293&0.03664&15.02&0.26&1.50&\\
8&1217&0.03448&15.93&0.26&1.50&\\
\hline
9&1689&0.04786&11.62&0.39&1.54&\\
10&1457&0.04128&13.39&0.39&1.51&\\
11&1299&0.03680&14.96&0.40&1.48&\\
12&1223&0.03465&15.86&0.40&1.50&\\
\hline
\end{tabular}
\end{table}

\begin{figure}[h!]
\centering
\includegraphics[width=16 cm]{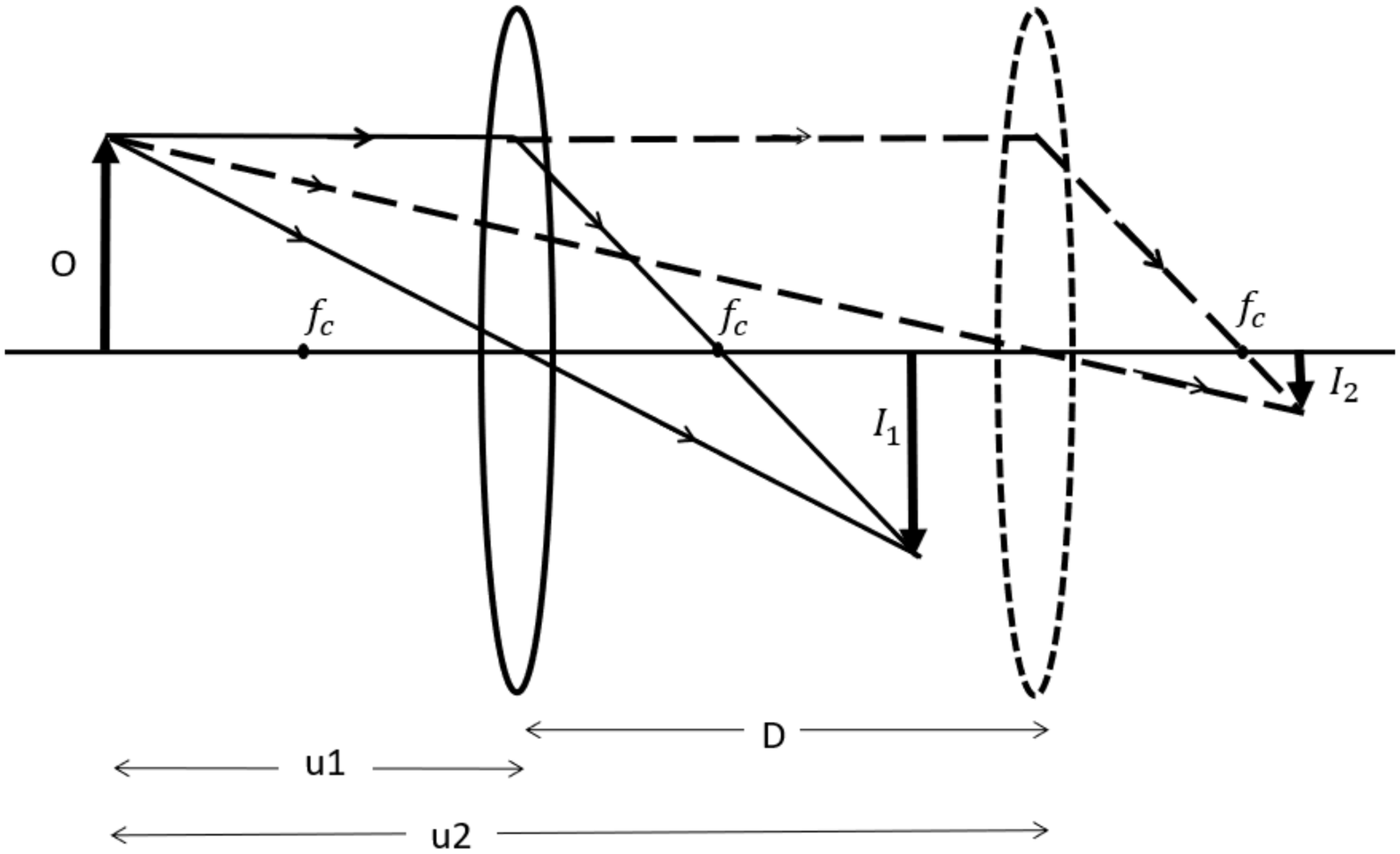}
Fig.1 {Image formation at the two positions of the lens}
\end{figure}

\newpage
\begin{center}
References
\end{center}
\begin{enumerate}
\item Wang J and Sun W 2019 Measuring the focal length of a camera lens in a smart-phone with a ruler Phys.Teach. \textbf{57} 54 
\item Girot A, Goy N-A, Vilquin A, and Delabre U 2020 Studying ray optics with a smartphone Phys.Teach. \textbf{58} 133
\item Sarkar S,Pal S, and Chakrabarti S 2021 Paper submitted to Phys.Teach.(AAPT)
\item Nassar A 1994 Apparent Depth Phys.Teach. \textbf{32} 526 
\item Newburg et al. 2000 Using the small-angle approximation to measure the index of refraction of water Phys.Teach. \textbf{38} 478
\item Baki S B O and Mansur Y  2019 Depth-imaging Analysis for Refractive Index Measurement Journal of Multidisciplinary Engineering Science Technique \textbf  {6(12)}    97, Special Issue
\item  Goehl J F 2003 Previous Apparent-Depth Papers Phys. Teach. \textbf {41} 4
\item Palmer G and Clark T A 1976 Determination of refractive index of glass Phys. Educ. \textbf {11(6)} 424
\item Lombardi S et al 2010 Measuring variable refractive indices using digital photos Phys. Educ. \textbf {45} 83 
\item https://www.metadata2go.com/
\item Atkins P and Paula J d 2002 Atkins' Physical Chemistry 7th edn (Oxford University Press) 1100
\item Tan C-Y and Huang Y-X 2015 Dependence of Refractive index on concentration and temperature in Electrolytic solution, Polar Solution, Nonpolar solution, and Protein solution Journal of Chemical and Engineering data (JCED) Sept.1,2015, doi:10.1021/acs.jced.5b00018; Sept.2, 2015 http://pubs.acs.org

\end{enumerate}

\end{document}